\documentclass[12pt]{article}
\usepackage{cite}
\usepackage{amssymb}
\usepackage{amsmath}

\newcommand{\bmat}{\left(\begin{array}}
\newcommand{\emat}{\end{array}\right)}

\def\Deq#1{\mbox{$D$=#1}}
\def\Neq#1{\mbox{$N$=#1}}

\def\yzero{\smash{\hbox{$y\kern-4pt\raise1pt\hbox{${}^\circ$}$}}}

\def\a{\alpha}
\def\b{\beta}
\def\g{\gamma}
\def\G{\Gamma}
\def\d{\delta}
\def\ut{\frac{1}{3}}
\def\um{\frac{1}{2}}
\def\t{\theta}
\def\Om{\Omega}

\def\si{\sigma}

\def\-{\hphantom{-}}

\def\s2{\frac{1}{2}}

\def\oh{\frac{1}{2}}
\def\ut{\frac{1}{3}}
\def\IT{\bf T}
\def\beq{\begin{equation}}
\def\eeq{\end{equation}}
\def\beqa{\begin{eqnarray}}
\def\eeqa{\end{eqnarray}}
\def\tr{{\rm tr \,}}
\def\Tr{{\rm Tr \,}}

\def\IF{\relax{\rm I\kern-.18em F}}
\def\II{\relax{\rm I\kern-.18em I}}
\def\IP{\relax{\rm I\kern-.18em P}}
\def\IC{\relax\hbox{\kern.25em$\inbar\kern-.3em{\rm C}$}}
\def\IR{\relax{\rm I\kern-.18em R}}

\def\Dsl{\,\raise.15ex\hbox{/}\mkern-13.5mu D} 
\def\IZ{\bf Z}
\def\IC{\bf C}

\newcommand{\drawsquare}[2]{\hbox{%
\rule{#2pt}{#1pt}\hskip-#2pt
\rule{#1pt}{#2pt}\hskip-#1pt
\rule[#1pt]{#1pt}{#2pt}}\rule[#1pt]{#2pt}{#2pt}\hskip-#2pt
\rule{#2pt}{#1pt}}
\newcommand{\Ysymm}{\raisebox{-.5pt}{\drawsquare{6.5}{0.4}}\hskip-0.4pt%
        \raisebox{-.5pt}{\drawsquare{6.5}{0.4}}}
\newcommand{\Yasymm}{\raisebox{-3.5pt}{\drawsquare{6.5}{0.4}}\hskip-6.9pt%
        \raisebox{3pt}{\drawsquare{6.5}{0.4}}}

\catcode`\@=11
\newdimen\@rotdimen
\newbox\@rotbox
\def\@vspec#1{\special{ps:#1}}
\def\@rotstart#1{\@vspec{gsave currentpoint currentpoint translate
   #1 neg exch neg exch translate}}
\def\@rotfinish{\@vspec{currentpoint grestore moveto}}
\def\@rotr#1{\@rotdimen=\ht#1\advance\@rotdimen by\dp#1%
   \hbox to\@rotdimen{\hskip\ht#1\vbox to\wd#1{\@rotstart{90 rotate}%
   \box#1\vss}\hss}\@rotfinish}
\def\@rotl#1{\@rotdimen=\ht#1\advance\@rotdimen by\dp#1%
   \hbox to\@rotdimen{\vbox to\wd#1{\vskip\wd#1\@rotstart{270 rotate}%
   \box#1\vss}\hss}\@rotfinish}%
\def\@rotu#1{\@rotdimen=\ht#1\advance\@rotdimen by\dp#1%
   \hbox to\wd#1{\hskip\wd#1\vbox to\@rotdimen{\vskip\@rotdimen
   \@rotstart{-1 dup scale}\box#1\vss}\hss}\@rotfinish}%
\def\@rotf#1{\hbox to\wd#1{\hskip\wd#1\@rotstart{-1 1 scale}%
   \box#1\hss}\@rotfinish}%
\def\rotate{\@ifnextchar[{\@rotate}{\@rotate[l]}}
\def\@rotate[#1]#2{\setbox\@rotbox=\hbox{#2}\@nameuse{@rot#1}\@rotbox}
\catcode`\@=12

\topmargin
-1.5cm
\textwidth
15.5cm
\textheight
23.5cm
\oddsidemargin
0.7cm
\evensidemargin
1.2cm
\begin{document}
\makeatletter
\@addtoreset{equation}{section}
\makeatother
\renewcommand{\theequation}{\thesection.\arabic{equation}}
\pagestyle{empty}
\rightline{CAB-IB/2901204} \rightline{\tt hep-th/0403262}
\vspace{0.5cm}
\begin{center}
{\LARGE {Particle models from orientifolds at \\ Gepner-orbifold points \\[10mm]
} }{\large {G. Aldazabal $^{1,2}$, E. C. Andr\'es $^1$, J. E. Juknevich $^1 $
\\[2mm]
} }{\small {\ $^1$ Instituto Balseiro, CNEA, Centro At{\'o}mico Bariloche,\\[%
-0.3em]
8400 S.C. de Bariloche, and \ $^2$CONICET, Argentina.\\[9mm]
} {\bf Abstract} \\[7mm]
}
\begin{minipage}[h]{14.0cm}
We consider configurations of stacks of orientifold planes and D-branes wrapped on a non trivial internal space of  the structure ${( Gepner \,\, model)^{c=3n} \times \IT^{2(3-n)}}/\IZ_N$, for $n=1,2,3$.
By performing simple moddings by discrete symmetries of Gepner models at orientifold points, consistent with a $\IZ_N$ orbifold action, we show that  projection on D-brane configurations can be  achieved, generically leading to  chiral gauge theories.
Either supersymmetric or non-supersymmetric (tachyon free)  models can be obtained.
We illustrate the procedure through some explicit examples.
\end{minipage}
\end{center}
\newpage \setcounter{page}{1} \pagestyle{plain}
%
\setcounter{footnote}{0}
\section{Introduction}
The identification of D-branes \cite{pol}, as an essential ingredient of the string theory structure,
lies at the centre of the radical change that lead to the present conception of string theory.
{}From a phenomenological approach, the fact that gauge interactions are confined on brane world volumes, has opened new perspectives for understanding the ways in which the Standard Model of fundamental interactions can be embedded in string theory.
A key issue in this context is the origin of chirality. It is known that either branes intersecting at angles [\citen{bachas,bdl,bgkl,aads,afirus,bbkl,uinter, imr,fhs,csu,csu2}]
(or equivalently in the presence of turned on fluxes) or branes stuck at singularities [\citen{dm,kaku,abpss,afiv,aiqu,aa}]
(or a combination of both \cite{afirus}) can lead to chiral fermions\footnote{An alternative proposal was made in \cite{fluxes} where turned on  RR and NS fluxes  could require chiral fermions}.
Concrete constructions of particle models on D-brane worlds  have been performed, mainly,  in the presence of toroidal  like background compactifications of Type II A/B string theory. In particular, in the context of Type II theory orientifolds D-branes are required to provide the RR charges to cancel orientifold plane ones.
Toroidal like compactifications are, generically, easy to handle and allow for a simple description in geometrical terms. Nevertheless, the possible space of string backgrounds is much more richer and deserves further investigation.
Relevant steps towards the understanding of the algebraic and geometrical interpretation of D-branes on Calabi-Yau manifolds have been performed (see, for instance [\citen{fuchs2, huis, ABbranes,dgjt,goma,mt}]).
Particularly appealing, in a first step  approach towards generic  non flat background descriptions, are the, so called, $N=2$ coset theories, where the internal space can be described  in terms of solvable $N=2$ conformal theories \cite{gepner,ks}.
In Gepner models, on  which we concentrate in this article, such  internal sector is spanned by a tensor product of $N=2$ minimal conformal theories. Each minimal model is characterized by an affine level  $k$ implying a central charge $c=\frac{3k}{2+k}$. Blocks central charges must add up to total internal charge  $c_{int}=9$ in order to ensure conformal anomaly cancellation.

 Extensive studies on  such backgrounds have been performed in perturbative heterotic string  and a growing understanding of type II orientifold theories on Gepner points has been  acquired more recently [\citen{gepnerori,aaln,blumen,bw,bhhw,dhs,bw2}].

{}From a phenomenological perspective two  \textit{minimum }non trivial requirements for a given string compactification to be acceptable are:\\
i.The gauge group must be big enough, in order to fit the standard model.\\
ii.Chiral fermions should be present.

Gepner orientifold constructions tend to lead to rank reduction,  due to (see \cite{bianchi} and references therein) discrete NS B field turned on (also understandable from the presence of different kinds of orientifold planes ~\cite{witten}). For instance, the so called ($D=4$) $(k=1)^9$ model, a parent $ Z_3$ orbifold, possesses rank 2 gauge groups. Presence of higher rank, orthogonal groups, was first advanced in \cite{blumen}, where examples were given, and further analysed in \cite{aaln}(see also \cite{blumen,bw,bhhw,dhs}). Actually, high values of $k$ do generically lead to several tadpole equations which allow for higher ranks.

The only groups present in orientifold models,  with complex representations (required for chirality), are unitary groups. However, in Gepner orientifolds, unitary groups do not always appear. For instance, for odd $k$ level and diagonal invariant couplings (i.e.  models with B type branes), in arbitrary dimensions, it is known (see i.e.\cite{bw} and index formula below) that, only orthogonal or symplectic groups can appear. On the other hand, in explicit simple model examples, where some minimal blocks are coupled through charge conjugate invariant,  unitary groups have been found \cite{aaln}.
However, going beyond diagonal invariants, for instance by using the charge conjugate invariant, generically leads to a high number of tadpole equations which can be very difficult to tackle in concrete computations. Promising models have been constructed in \cite{bw} by projecting by simple current symmetries  leading to a reduction in the number of tadpole equations.
For even values of the level $k$ unitary groups are known to appear (see \cite{abpss,aaln,bhhw,dhs}. Such cases have some extra complications due to the presence of long and short orbits.

In \cite{aaln} a general description of modding by minimal model phase symmetries in Gepner orientifold was provided.

In the present article  we show that, by embedding these phase symmetry transformations of internal blocks into a $Z_N$ action on Chan Paton factors, in a very similar way like  an orbifold action on internal coordinates is embedded in gauge degrees of freedom,  unitary groups with chiral matter can be easily obtained.
In practice, it proves useful to start with diagonal invariant couplings which, as mentioned, are much easier to handle since they lead to a minimum number of tadpole cancellation equations. Thus, once solutions for such models have been found, leading to orthogonal or symplectic gauge theories for  $k$ odd, chiral theories are obtained by phase projecting.
{}From the point of view of tadpole cancellation, modding leads to an increase in the number of equations to be solved due to the appearance of massless twisted RR states in the transverse channel. Interestingly enough, by knowing the form of a definite modding, a certain  control on such number can be maintained \footnote{in a somewhat opposite way to \cite{bw} where projections are used to reduce the number of equations}.
Moreover, tadpole equations for modded theory can be rather easily obtained from non-modded one.

It appears natural to think in terms of internal spaces where some complex planes are fulfilled by  a Gepner model while the others are compactified on a torus. In such a scheme, stacks of orientifold planes and D-branes will wrap on non trivial cycles on the Gepner planes whereas parallel branes are expected to appear on the $C^{(3-n)}$ planes.
We consider such situations in this paper.
In particular, we show that, phase modding on Gepner sector can be accordingly accompanied by an orbifold projection on the torus  leading again to chiral theories. Schematically the internal sector would be described by ${( Gepner \,\, model)^{c=3n} \times \IT^{2(3-n)}}/ \IZ_N$, for $n=1,2,3$.
A rather similar idea was presented in \cite{afirus} with D-branes intersecting at angles in the $C^n$ plane but parallel on the remaining internal directions.
An advantage of this approach is that,  gauge group rank can be reduced, due to presence of Gepner term, but remain  big enough in order to lead to phenomenologically interesting constructions.

The paper is organized as follows. Section 2 contains a generic  introduction to Type IIB orientifold ideas. In Section 3 orbifold  and Gepner like internal spaces, as well as hybrid orbifold-Gepner cases, are briefly discussed. Modding by phase symmetries is introduced and the construction of the corresponding supersymmetric projected characters is shown.
Notation and general results closely follow those of reference \cite{aaln}.
The open string sector is discussed in section 4. Special emphasis is given to the way in which phase moddings are embedded as Chan-Paton factor twists. An index formula is presented. Tadpole cancellation is addressed in section 5.
In section 6 we show, through some simple examples, how chiral models can be easily obtained in both supersymmetric and non supersymmetric models. Details are left to the Appendix.

\textit{Note}: While this work was under completion we become aware of the nice results of Ref.\cite{dhs}. Even if the two approaches are different there is still some overlap.

\section{Type II orientifolds}
In this section we   briefly review the basic steps implied in the construction of orientifold models.
Essentially an orientifold model is obtained by dividing out the orientation reversal symmetry of Type II string theory.
Schematically, Type IIB torus partition function is defined as
\begin{equation}\label{2bpf}
{\cal Z}_{T} (\tau, {\bar \tau})=   \sum_{a,b} \chi_a(\tau )
{\cal N}^{ab} {\bar \chi}_b({\bar \tau })
\end{equation}
where the characters $\chi_a(\tau )= \Tr_{{\cal H}a}
q^{L_0-\frac{c}{24} }$, with $q= e^{2i\pi \tau}$, span a representation
of the modular group of the torus
generated by {\sf S}: $\tau \to -\frac1{\tau}$ and {\sf T}: $\tau
\to\tau+1$ transformations.
${\cal H}_a$ is the Hilbert space of a conformal field theory with central
charge
$c=15$
generated from
a conformal primary state $\phi _a$
(similarly for the right moving algebra).
In particular $\chi_a( -\frac1{\tau} )= S_{aa'}\chi_{a'}(\tau )$
and modular invariance requires
$ S {\cal N} S^{-1}= {\cal N}$(for left -right symmetric theories ${\cal N}^{ab}={\cal N}^{ba}$).
Generically, the characters can be split into a spacetime piece,
contributing with
$c_{st}= {\bar c }_{st}= {\frac32} D$ and an internal sector with
$c_{int} = {\bar c}_{int}= {\frac32}(10-D)$.

Let $\Omega $ be the reversing order (orientifolding)  operator permuting
right and left movers.
Modding by  order reversal symmetry is then
implemented by introducing the projection operator
$\frac12 (1+\Omega)$ into the torus partition function.
The resulting vacuum amplitude reads
\begin{equation}\label{clomega}
{\cal Z}_{\Omega} (\tau, {\bar \tau})= {\cal Z}_{T} (\tau, {\bar \tau})
+ {\cal Z}_{K} (\tau- {\bar \tau}) .
\end{equation}
 The first contribution is just the symmetrization
(or anti-symmetrization in
case states anticommute) of left and right sector contributions
indicating that two states differing in a left-right ordering must be
counted once.
The second term is the
Klein bottle contribution and takes into account states that are
exactly the same in both sectors.
In such case, the operator
$e^{2i \pi \tau {L_0}}
e^{-2i \pi {\bar \tau}{{\bar L}_0}}$,
when acting on the same states, becomes
$e^{2i \pi 2it_K {L_0}}$ with $\tau -{\bar {\tau}}= 2it_K$ and thus
\begin{equation}\label{kbd}
{\cal Z}_{K} (2it_K) = \frac 12 \sum_{a} {\cal K}^{a}  \chi_a(2it_K) ,
\end{equation}
where $|{\cal K}^{a}|={\cal N}^{aa}$.
The Klein bottle
amplitude in the {\it transverse channel} is obtained by performing
an {\sf S} modular transformation
such that
\begin{equation}\label{kbo}
\tilde {\cal Z}_{K} (i l)= \frac 12 \sum_{a} O^2_a  \chi_a(il )
\end{equation}
with $l=\frac1{2t_K}$ and
\beq
O^2_a= 2^D {\cal K}^{b}S_{ba}
\eeq
This notation for the closed channel coefficients highlights the fact
that the Klein bottle
 transverse channel represents a closed string propagating
between two
crosscaps (orientifold planes) which act  like boundaries. When integrated over
the tube length, such amplitude leads, for massless states, to tadpole like divergences. In particular, for RR massless states such tadpoles must be cancelled for the theory to be consistent.
Notice that, for such fields, $O_a$ represents the charge of the orientifold plane (crosscap) under them and, therefore, inclusion of an open string sector with D-branes carrying $-O_a$ RR
 charge provides
a way for having a consistent theory \cite{pol,gp, pchj} with net
vanishing charge.
An open string cylinder amplitude,
representing strings propagating between two D-branes, and
a M\"obius strip amplitude with strings propagating between orientifold
planes and D-branes must be included.
In the long tube limit the sum of the contributions from the
Klein bottle, cylinder and M\"obius strip
in the transverse channel must then  factorize as
\begin{equation}
\tilde {\cal Z}_{K} (il)
+\tilde {\cal Z}_M (il) +
\tilde {\cal Z}_C (i l)\to
\sum_{a} (O_a +D_a)^2 \frac1{{m_a}^2}= \sum_{a} (O_a^2
+2 O_a D_a+ D_a^2) \frac1{{m_a}^2}
\label{factori}
\end{equation}
where ${m_a}$ is the mass of the state in  $\chi_a$.
For massless RR fields  $ D_a$ is the D-brane RR charge
and  absence of divergences requires
\begin{equation}
O_a +D_a=0 .
\label{tadpolecg}
\end{equation}
Cylinder amplitude in the direct channel should
read
\begin{equation}\label{cild}
{\cal Z}_{C} (it_{C}) =\oh  \sum_{a} {\cal C}_{a}  \chi_a(it_{C} ) ,
\end{equation}
where
\beq
{\cal C}_{a} = C_{jka} n_j n_k
\eeq
represents the multiplicity of states contained in $ \chi_a(it )$ and
$n_j$, $n_k$ are Chan-Paton multiplicities.
Namely, open sector states are of the form
\begin{equation}
|\Phi_k ; i,j\rangle \lambda^k_{ji}
\end{equation}
where $\Phi_k$ is a world sheet  conformal field and $j$ ($i$) label the 
type of branes where the string endpoints
must be attached. $n_j$ is the number of branes on each stack while $\lambda^k_{ji}$ represent the corresponding wave functions in this brane basis expansion.
 $C_{ija}$ are positive integers (actually ${C_{ija}}= 0,1,2$) 
 generated when trace over open states $|\Phi_k ; i,j\rangle$ is computed.
When rewriting \ref{cild} in the  transverse channel we recover $\tilde {\cal Z}_C (i l)$ in \ref{factori} where 
with  $ D_a=D_{ja} n_j $ and
\beq
(D_{ja} n_j)^2= {\cal C}_{b} S_{ba}= C_{jkb} n_j  n_k S_{ba}
\label{da}
\eeq
The  M\"obius strip amplitude is constructed in a similar way. However, since characters $\Tr_{{\cal H}_a}(e^{\pi
it(L_{0}-\frac{%
c}{24})}\Omega )=\chi_a (it_{M}+\frac{1}{2})$ are non-real, it proves convenient to work in terms of the real  ``hatted"
${\hat
\chi}_a(il+\oh )=e^{i\pi
(h_a-c/24)} \chi_a (it_{M}+\frac{1}{2})$ characters.
Thus, MS amplitude in the direct channel takes the form
\begin{equation}\label{msd}
{\cal Z}_{M} (it_M) =\oh  \sum_{a} {\cal M}_{a}  {\hat \chi}_a(it_{M}+\oh
)
\end{equation}
where now
\beq
{\cal M}_{a} = M_{ja}n_j
\eeq
are integer numbers.
The characters in the direct and transverse channels of the M\"obius strip
are related by the transformation \cite{bs1}
{\sf P}: $it_M+\frac{1}{2}
\to \frac{i}{4t_M}+\frac{1}{2}$
generated
from the modular transformations {\sf S} and {\sf T} as
${\rm \sf P}={\rm {\sf TST}}^{2}{\sf S}$.
Thus, we can rewrite the transverse channel in \ref{factori} representing a
closed string propagating between a D-brane and an orientifold plane from where we read
\beq
O_a (D_{ja} n_j)= 2^{\frac D2}{\cal M}_{b} P_{ba}= 2^{\frac D2}M_{jb}n_j
P_{ba}
\eeq

In principle, once the Klein bottle partition function is obtained from the left-right
symmetric
type IIB torus partition function, a full consistent open string theory can be constructed by ensuring factorization, massless RR tadpole cancellation, and consistency restrictions on the integer coefficients $C_{jia}$ and
$M_{ja}$. Certainly, such steps can be more or less cumbersome depending on the type of models considered.
\section{$D=4$ Type IIB orientifolds on  Gepner models and orbifolds}
\medskip
In $D=4$ dimensions each moving sector of Type IIB theory  is described by a conformal theory of total central charge $c_{tot}=c_{st}+c_{int}=12$, where $c_{st}=3$ and $c_{int}=9$ are the central charges corresponding to space-time and internal (six dimensional) sectors, respectively.
In a toroidal like compactification six free bosonic and fermionic fields, each one contributing with 1 and $1/2$ units to central charge respectively, curl up the extra six dimensions in order to provide a consistent theory. Such compactifications are generically non chiral,  since too many supersymmetries are preserved.
If  orbifold like singularities are present some or all of supersymmetry generators are projected out.
For instance, consider  $\IZ_N$ orbifold action performed by the generator $\theta $ such that
\begin{equation}
\theta^x Y_i  = {\rm e}^{2i\pi xv_i}Y_i
\label{thdi}
\end{equation}
with $x$ an integer number and where $Y_I$, $I=1,2,3$ are complex bosonic coordinates parameterizing the $\IT^{6}$ internal torus. The twist vector $v= (v_1, v_2, v_3)$ encodes the orbifold action on each complex plane.
Thus, for instance, for untwisted massless Left (or Right) Ramond states of the form
$| \si_0 \si_1 \si_2 \si_3 \rangle$
with $\si_0, \si_i = \pm \oh$, we have
\beq
\theta^x |\si _0 \si _1 \si _2 \si _3 \rangle = {\rm e}^{2i\pi x v\cdot \si}
|\si_0 \si_1 \si_2 \si_3 \rangle
\label{twistR}
\eeq
The invariance condition
\begin{equation}
\si_Iv_I \in \mathbb{Z}
\label{orbieigenvalue}
\end{equation}
projects  some of the fermionic states out and therefore reduces  the number of supersymmetries.

In particular, the condition $\pm v_1 \pm v_2 \pm v_3= 0$ ensures that there is a
gravitino in both the NS-R and R-NS type IIB untwisted sectors.
Projection under $\Om$, produces the closed sector of the orientifold and then leads to \Neq1, \Deq4 supersymmetry.
Partition function can be found, for instance,  in Ref.\cite{afiv}.

Gepner models \cite{gepner} offer  an alternative in which  supersymmetric
string vacua are provided in terms of an explicit algebraic construction. The internal sector is given by a tensor product of $r$
copies of N=2 superconformal minimal models with levels $k_j$, $j=1,...,r$
and central charge
\begin{equation}
c = \frac{3k}{k+2} \quad , \quad k=1,2,...  \label{mm}
\end{equation}
with total central charge $\sum _{j=1}^r c_{k_j}^{int}=9$.
Spacetime supersymmetry and modular invariance are implemented by keeping in
the spectrum only states for which the total $U(1)$ charge is an odd
integer.
More explicitely,  $N=2$ minimal models unitary representations, are
encoded in primary fields   labelled by three integers $(l,q,s)$ such that $l=0,1,...,k$; $%
l+q+s=0$ mod 2. They belong to the NS or R sector when $l+q$ is
even or
odd respectively. The conformal dimensions and charges of the highest
weight
states are given by
\begin{eqnarray}
\Delta_{l,q,s} = \frac{l(l+2)-q^2}{4(k+2)} + \frac {s^2}{8} \quad {\rm mod}
~1  \label{peso} \\
Q_{l,q,s} = -\frac{q}{k+2}+\frac s2 \quad {\rm mod}~ 2 .  \label{carga}
\end{eqnarray}
Two representations labelled by $(l^{\prime},q^{\prime},s^{\prime})$ and $%
(l,q,s)$ are equivalent, $i.e.$ they correspond to the same state, if
\begin{equation}
l^{\prime}=l  \quad , \quad
q^{\prime}= q ~{\rm mod} ~ 2(k+2)\quad , \quad s^{\prime}=s ~ {\rm mod}~4
\label{id1}
\end{equation}
or
\begin{equation}
l^{\prime}=k-l \quad , \quad q^{\prime}=q+k+2 \quad , \quad s^{\prime}=s+2
\label{id2}
\end{equation}
The exact conformal dimension and
charge of the highest weight state in the representation $(l,q,s)$ are
obtained from equations (\ref{peso}) and (\ref{carga}) using the
identifications above to bring $(l,q,s)$ to the $standard ~ range$ given by
\begin{equation}
l=0,1,...,k \quad ; \quad |q-s| \le l \quad ; \quad l+q+s=0 ~ {\rm mod} ~ 2
\end{equation}
and $|s|$ is the minimum value among those in (\ref{id1}) and (\ref{id2}).

The partition function of the minimal models on the torus can be written in
terms of the characters of the irreducible representations as
\begin{equation}
{\cal Z}^{(m.m.)}_T(\tau) = \sum_{(l, q, s),(\bar l, \bar q, \bar s)} {\cal N%
}_ {(l, q, s),(\bar l, \bar q, \bar s)} \chi_{(l, q, s)}(\tau,0) \chi_{(\bar
l, \bar q, \bar s)}^*(\bar\tau,0)
\end{equation}
where the coefficients ${\cal N}_{(l, q, s),(\bar l, \bar q, \bar s)}$ are
non negative integer numbers which count the number of times the irreducible
representation $(l, q, s)\otimes(\bar l, \bar q, \bar s)$ is contained in $%
{\cal H}$. The existence of a unique ground state requires ${\cal N}%
_{(0,0,0),(0,0,0)} = 1$.
The characters in the sector ${\cal H}_{(l,q,s)}$ are given by
\begin{equation}
\chi_{(l, q, s)}(\tau,z) = \Tr_{{\cal H}_{(l, q, s)}} \left ( e^{2\pi i
\tau (L_0 - \frac c{24})} e^{2\pi i z J_0}\right )  \label{cardef}
\end{equation}
and it proves useful to define the combination
\beq
\chi _{l,q}(\tau ,z)=\chi _{(l,q,s)}(\tau ,z)+\chi
_{(l,q,s+2)}(\tau ,z)
\label{carss}
\eeq
Characters of  Gepner model are  obtained by tensoring the space time  and the $r$ internal minimal models characters with the constraint on  total $U(1)$ charge to be odd, in order to ensure one supersymmetry. Namely,
\begin{equation}
Q_{tot}= Q_{\nu}+\sum _{j=1}^rQ_{l_j,q_j,s_j} \in  2\mathbb{Z}+1
\label{oddcharge}
\end{equation}
where $\nu=1,-1,0,2$ refers to spinor, conjugate spinor, scalar and vector representations, respectively.
Thus, by defining
\begin{equation}
\chi_{\vec \alpha}(\tau, z) =\{
[\chi_{\nu}(\tau,
z)]^d \chi_{\alpha_1}(\tau, z) \chi_{\alpha_2}(\tau, z)\dots \chi_{\alpha_r}(\tau, z)\}
\label{alpha}
\end{equation}
with
\begin{equation}
\vec \alpha=(\alpha_0,\alpha_1,\dots,\alpha_r)    \, \, \, \, \, \,  \, \, \alpha_j=(l_j,q_j)
\end{equation}
where $[\chi_{\nu}(\tau, z)]^d$ is the $D$ dimensional spacetime character
with $d=\frac {(D-2)}2$.

A  \textit{supersymmetric}
character,  given by
\beqa
& & \chi^{susy}_{\vec\alpha}(\tau, z) = \sum_{n=0}^{2m-1} (-1)^{n}
\chi_{\vec \alpha^{(n)}} (\tau, z)=  \label{susych} \\
& &= \frac{1}{2m}\sum\limits_{n,p\;{\rm mod\;2m}}(-1)^{n+p}
e^{2\pi i(n^{2}\frac{c}{24}\tau +n\frac{c}{6}z)}
\left [\chi_0(\tau
,z+\frac n2 \tau+\frac p2)
\right ]^{d}
\prod\limits_{i=1}^{r}\chi _{l_{i},q_{i}}(\tau ,z+\frac{n%
}{2}\tau +\frac{p}{2}) \nonumber
\label{chisusy}
\eeqa
can be built (  $c=12$ here).
NS or R sectors are obtained when
summing over even or odd $n$ respectively, whereas periodic ($+$) or
antiperiodic ($-$) characters arise when summing over even or odd $p$, respectively.
Odd charge condition \ref{oddcharge} is ensured by the sum over $p$.
The  \textit{susy} character does transform as the non-susy one and, therefore, full modular invariant partition function is obtained by just coupling the right and left sectors as
\begin{equation}
{\cal Z}_T(\tau, \bar \tau) = \sum_ {\vec \alpha ;
\vec {\bar \alpha}} {\cal N}_{\vec \alpha ; \vec {\bar \alpha}}
\chi^{susy}_{\vec\alpha}(\tau,0) \chi^{susy ~ *}_{\vec {\bar
\alpha}}(\bar\tau,0)
\end{equation}

${\cal N}_{\vec\alpha , \vec{\bar\alpha}}$ are positive integer
coefficients obtained from the product $\prod_{i=1}^r{\cal N}_{\alpha_i;\bar
\alpha_i}$ of the individual minimal models.
An integration over $\tau$, with the appropriate measure, must then be performed.

Closed sector is obtained by keeping  $\Omega $ invariant states while open sector  can then be easily written down,  as linear combinations of these  explicitly  supersymmetric characters.

\subsection{Phase moddings}

Gepner models posses a phase symmetry group $G=\otimes_a  Z_{M_a}$, associated to phase transformations of primary fields
\beq
\Phi_{l,q,s}
\to e^{-2i \pi \gamma {\frac{q}{m} }}\Phi_{l,q,s}
\label{phsym}
\eeq with $\gamma \in Z$
in each block. Thus, the full phase transformation can be encoded  into an $r$ dimensional
vector
\beq
{{\vec \Gamma  }}^{ a} = ( { \gamma_1}^{a }  ,{\gamma_2}^{a }, \dots,{\gamma_r}^{ a})
\label{gamma}
\eeq
where $M_a$ is the least integer such that
\beq
M_a { \gamma_i}^{a }= 0 \quad \mod \, (k_i+2)
\eeq
and $a $ labels one of the different, inequivalent, phase transformations.
Moddings by such symmetries can be easily implemented \cite{aaln} by replacing character $\chi_{{\vec l},{\vec q }}(\tau )\rightarrow \chi^{G}_{{\vec l},{\vec q }}(\tau )$ in  \ref{chisusy} where  the projected character reads
\beq
 \chi^{G}_{{\vec l},{\vec q }}(\tau )=\frac{1}{M}  \sum_{x,y} \chi^G_{{\vec l},{\vec q }}(\tau,x,y )
\eeq
We have defined the character in sector $(x,y)$ as
\beq
\chi^G_{{\vec l},{\vec q }}(\tau,x,y )=
e^{-2i \pi {x \frac{\gamma_i}{m}(q_i+ \gamma_i y}) }
\chi_{{\vec l},{\vec q }+2{\vec \gamma}y}(\tau)
\label{modchi}
\eeq
where $\vec l$, $\vec q$ are $r$-component vectors with entries $l_i$,
$q_i$ respectively.
The  projection conditions on each twisted closed sector, $y =0,\dots,M_a-1$, are

\begin{equation}\label{twistedpro}
   \sum_{i=1}^r   \frac1m_i { \gamma_i^{a }
(q_i+y \gamma_i^{a } )}
\in {\mathbb Z}
\end{equation}
while supersymmetry imposes the further constraint
${ \gamma_i}^{a }$,
\begin{equation}\label{susyconst}
 \sum_{i=1}^r  \frac1m_i \gamma_i^{a }
\in {\mathbb Z}
\end{equation}
(This is the usual $2\beta_0\cdot\Gamma
\in {\mathbb Z}
$ condition of
\cite{gepner}) ensuring integer  total charge).
A key point in the construction of the projected characters is that they  transform as the original ones under modular transformations. Namely,
\begin{eqnarray}
S&:&\rightarrow \chi^G_{\vec\alpha}(-\frac{1}{\tau},x,y )= (-i \tau)^{-1}\sum_{\vec\beta}
S_{\vec\alpha,\vec\beta} \chi^G_{\vec\beta}(\tau,-y,x ) \label{matrizs}\\
T &:&\rightarrow \chi^G_{\vec\alpha}(\tau +1,x,y )=e^{i \pi(\Delta_{\a} - \frac{Q_{\a}}{2}- \frac{c}{24}) } \chi^G_{\vec\alpha}(\tau,x+y,x )\label{matrizt}
\end{eqnarray}
where $Q_{\a}= -\sum \frac{q_i}{m_i}$, $\Delta_{\a}= \sum \Delta_i$.
Notice that same steps can be repeated for right moving sector with characters now depending on $(\bar x, \bar y)$. Moreover, different moddings on right and left  moving sectors could be performed.

Interestingly enough, it is possible to think in terms of a kind of hybrid compactification where part of the internal sector is built up from a Gepner model while the rest corresponds to a toroidal like compactification.
More specifically, let us  assume that we start with a $c=3n$ $N=1$ Gepner model in $d=10-2n$ dimensions and that we further compactify $2(3-n)$ coordinates on a torus in order to obtain a  four dimensional model (with extended supersymmetry). A massless, lets say left sector state, would read
\begin{equation}
|r_0,r_1,\dots,r_{3-n},(l_i,q_i,s_i)_{i=1,...,r}\rangle               
\label{leftstate}
\end{equation}
where  $r_i$ $i=0,\dots,3-n$ are  $SO(2(3-n))$ weight vectors and, a generalized, GSO projection requires
\begin{equation}
\sum_{i=0}^{ 3-n} r_i-\sum_{j=1}^{ r}\oh s _j-\sum_{j=1}^{
  r}\frac{q_j}{m_j} \quad \in 2\mathbb{Z}+1
\end{equation}

If the toroidal sector has a symmetry $ \IZ_N$ (generically $\IZ_N \times \IZ_M$), the full internal sector will be invariant under the symmetry group
$ G=\IZ_N \otimes_a  Z_{M_a}$ with ${a=1,\dots,r}$.
Orbifolding by such a symmetry, the internal space would look like
$( Gepner\, \, model)^{c=3n} \times  \IT^{2(3-n)}/G$, for $n=1,2,3$
where orbifold action is encoded in phase  moddings  vectors $\Gamma$ of \ref{gamma} of the Gepner sector and eigenvalue vector $v$ \ref{orbieigenvalue} for the $\IZ_N$ orbifold action on the  internal tori. Both actions must be  performed simultaneously, in a compatible, modular invariant form and will lead to a reduction of supersymmetry.
Namely, consider  left fermion $Q_{\oh} $ associated to Susy generators. They are of the form shown in \ref{leftstate} with $r_0=1/2$. If $\theta$ realizes a $G$ twist then
\begin{equation}
\theta Q^A_{\mathbf {\oh}}\theta ^{-1} = e^{2\pi i(v_i  r_i - \frac{\g_i q_i}{m_i})} Q^A_{\mathbf {\oh}}.
\end{equation}
Therefore, for supersymmetry  to be preserved
\begin{equation}
v_i r_i - \frac{\g_i q_i}{m_i} \quad \in \mathbb{Z}   \label{forsusy}
\end{equation}
In particular, a  $ \IZ_M$ subgroup of $G$, whose action is  encoded in  twist eigenvalues  $(\G,v)$, can be chosen in order to keep just $N=1$ supersymmetries.

Thus, the $(x,y)$ sector of a $\Gamma $ modded  Gepner model is accompanied by a $(\theta ^x, \theta ^y)$ action on the internal torus. Calling  $\chi_{\a,x,y}(\tau)$ the partition contribution to such twisted $(x,y)$ sector  (and similarly for $(\bar{x},\bar{y})$) the partition function will formally read
\begin{equation}
Z_T = \frac{1}{M}\sum_{\a, \b  ,x , y, \bar x, \bar y} \tilde \chi(x, y,\bar x,\bar y) N_{\a\b}
e^{-2\pi i \frac{\G x}{m}(q + y \G)} \chi_{\a,x,y}(q)
e^{-2\pi i \frac{\G \bar x}{m}(q - \bar y \G)} \chi_{\b,\bar x, -\bar y}(\bar q)
\end{equation}
where $\tilde \chi(x, y,\bar x,\bar y)$
is the usual fixed point multiplicity (see \cite{imnq})  and
\begin{equation}
\chi_{\a,x,y}(\tau)=\left \{
\left [\frac{\theta\left[{0 \atop 0}\right]}{\eta^3} \right]^2
\prod_{i=1}^{3-n} e^{i\pi y v_i}\frac{\theta\left[{yv_i \atop {xv_i}}\right]}
{\theta\left[{\frac{1}{2} + yv_i \atop {\frac{1}{2}+ xv_i}}\right ]}
 \chi_{\a +2 \G y} \right \}_{susy}
\end{equation}
are supersymmetric characters $\chi_{\a,x,y}$ for left movers (or right).
The first terms in a $q, \bar q$ expansion read,$(q=e^{-2\pi t})$
\begin{eqnarray*}
Z(\tau) &=& \frac{1}{M} \sum N_{\a\b}\tilde \chi(x,y,\bar x, \bar y)\\
& & e^{2\pi i(r+yv)xv}e^{-2\pi i \frac{\G x}{m}(q + y \G)}
q^{\oh(r+yv)^2 + E_0( y) +\sum h_{\a,y} -\oh}(1+...) \\
& & e^{-2\pi i(\tilde r + \bar y v)\bar x v}e^{-2\pi i \frac{\G \bar x}{m}(\bar q - \bar y \G)}
{\bar q }^{\oh(\tilde r+\bar y v)^2 + E_0(\bar y) +\sum \bar h_{\b,\bar y} -\oh}(1+...)
\label{xypartfunct}
\end{eqnarray*}
where $E_0(y)= \sum_i \oh|yv_i|(1-|yv_i|)$ is the zero point energy and $h_{\a,y}$ is the conformal weight of the primary fields contained in $\chi_{\a,x,y}$.
Here $r, \tilde r$ are $SO(2n+2) $ weight vectors.
Thus, for a physical state twisted by $y$, encoded in the numbers $(r,\a)_y \equiv (r,l_i,q_i,s_i)_y$  we can read, for instance, the conditions to have  massless state \footnote{For massive states Gepner descendants and oscillators must be included.}
\begin{equation}
\oh(r+yv)^2 + E_0( y) +\sum h_{\a,y} -\oh =\oh(\tilde r+\bar y v)^2 + E_0(\bar y) +\sum \bar h_{\a,\bar y}-\oh=0
\end{equation}
with the projector onto invariant states given by
\begin{equation}
D(y,\bar y)= \frac{1}{M}\sum_{x,\bar x} \tilde \chi(x,y,\bar x ,\bar y)
e^{2\pi i(r+yv)xv}e^{-2\pi i \frac{\G x}{m}(q + y \G)}
e^{-2\pi i(\tilde r + \bar y v)\bar x v}e^{2\pi i \frac{ \G \bar x}{m}(\bar q + \bar y \G)}
\end{equation}
Let us return to the pure Gepner case with the aim to deduce
an expression for the Klein bottle amplitude.
A particular interesting case arises  when no modding is performed on
the right sector ($\bar \G=0$) in the {\em diagonal} invariant partition
function. In this case the partition function reads
\begin{equation}
Z_T = \sum_{\a}  \left (  \frac{1}{M} \sum_{x,y} e^{-2\pi i \frac{\G}{m}x(q + \G y)}
\chi_{\a + 2 \G y} \right ) \chi_{\a}^{*}
\end{equation}

 which, after summing over $x$, can be rewritten in the following way
\begin{equation}
Z_T = \sum_{\a} \left ( \sum_{y} \d \left ( \frac{\G}{m}(q + \G y) \right )
\chi_{\a + 2 \G y} \right ) \chi_{\a}^{*}.
\end{equation}
Modular invariance can be easily checked from \ref{matrizs}.

{}From this latter expression it follows that orbifolding
the CFT with respect to phase symmetries induces more general
invariant partition function besides the diagonal one.

Klein bottle amplitude is found by keeping identical right and left states. Thus  $y=0 \mod{ m/2}$. In particular, for odd $m$,  only $y=0$ states are allowed leading to
\begin{equation}
Z_K = \sum_{\a} \d \left ( \frac{\G q}{m} \right )
\chi_{\a } (2it)
\label{kbmodded}
\end{equation}

We notice that, when $m$ is odd, i.e. the case we are mainly considering in this article, there are no $y$ twisted sectors at all in the direct channel. $Z_M$ action on KB amplitude reduces to a projection onto $(\G)$ invariant states.
Notice, however, that $(x,0)$ sectors will go into $(0,x)$ twisted sectors in transverse channel and could lead to new contributions to tadpoles. By transforming the Klein bottle amplitude to the transverse channel, $(l=\frac{1}{t})$, previously
writing the constraint $\d \left ( \frac{\G q}{m} \right )$ as $\frac{1}{M}\sum e^{-2i\pi \G x\frac{q}{m}}$ we obtain
\begin{equation}
\tilde Z_K=\frac{1}{M}\sum_{x} \sum_{\a \b}2^D K_{\a}S_{\a\b} \chi_{\b+2\G x}(il)
\end{equation}
As mentioned, generically, massless RR fields will be present such that
$\tilde Z_K(il)$ will lead to undesired tadpole divergencies when integrated over $l$.
Therefore  D-branes amplitudes must be included in order to cancel such divergences.
Similar considerations apply to the hybrid model with the additional complication
that, in the present case, characters also depend on both x and y. For odd M, which is the case that  we are mainly studying, the Klein bottle amplitude reads
\begin{equation}
Z_K(2it)= \left \{ \frac{1}{M}\sum_{x,\a}
\left [\frac{\theta\left[{0\atop 0}\right]}{\eta^3} \right]^2
\prod_{i=1}^{3-n} \frac{\theta\left[{0 \atop {xv_i}}\right]}{\eta} \frac{-2 \sin{\pi x v_i }\eta}
{\theta\left[{\frac{1}{2} \atop {\frac{1}{2}+xv_i}}\right]} K_{\a} e^{-2i\pi \G x\frac{q}{m}} \chi_{\a} \right \}_{susy}
\label{kleinorb}
\end{equation}
which results from plugging the known partition function for orbifolds and Gepner models.
\section{Open sector}
Open string states  formally read
\begin{equation}
|\Phi_k ; i,j\rangle \lambda^k_{ji}
\end{equation}
where $\lambda^k $ encodes the gauge group representation in which the state $\Phi_k$ transforms. For instance, if the state $\Phi_0$ corresponds to gauge bosons, $ \lambda^0$ represents  gauge group $G$ generators \footnote{ Which generically will be a product of unitary, orthogonal and symplectic groups.}.

It proves useful \cite{afiv} to write down Chan Paton matrices in a
Cartan-Weyl basis where  generators organize into charged generators
$\lambda_a = E_a$, $a=1,\cdots, {\rm dim}\; G$, and Cartan generators
$\lambda_I = H_I$, $I=1,\cdots, {\rm rank}\; G$.

In such basis, information about gauge group and matter representations can be encoded into the corresponding root and weight vectors.
The allowed $\lambda^k$ are determined  by consistency of the full open string theory, ensuring factorization, tadpole cancellation and classical gauge groups with two indices representations.

Let us assume that such Chan-Paton factors have already  been determined and that we further act on string states with a generator $\theta$ of a $Z_M$ symmetry group. Such action which manifests as a phase $\d_k$ on world sheet field $\Phi_k$ should, in principle, be accompanied by corresponding representation of group action such that
\begin{eqnarray*}
\hat \theta|\Phi_k ; i,j\rangle \lambda_{ji}& = &\g_{ii'}|\hat \theta\Phi_k ; i',j'\rangle \g_{j'j}\lambda_{ji} \\
& = &e^{2\pi i \d_k}(\g^{-1}\lambda\g)_{j'i'}|\Phi_k ; i',j'\rangle
\end{eqnarray*}
Therefore, invariance under such action requires
\begin{equation}
e^{2\pi i \d_k} \g^{-1}\lambda ^k \g = \lambda ^k
\label{thetaprojection}
\end{equation}

For odd $M$ and the partition function given in \ref{kbmodded} we know that $\IZ_M$ twist reduces to a projection with no $y$ twisted sector.
Thus, for a generic hybrid case where  the internal sector of the form ${( Gepner \,\, model)^{c=3n} \times \IT^{2(3-n)}}/ \IZ_M$, for ($n=1,2,3$),  only  open string   $(v,\G)$ invariant states will remain in the spectrum with Chan-Paton factors satisfying the above equation with
\begin{equation}
\delta _k=(v.r -\frac{\G .q}{m} )
\end{equation}
By following the same steps as in Ref. \cite{afiv}, we can represent $Z_M$ Chan-Paton twist in terms of Cartan generators as $\g=e^{2\pi i VH}$
where $V$ is a ``shift"  eigenvalues vector of the generic form
\begin{equation}
V=\frac{1}{M}(0^{N_0},1^{N_1},\dots,(M-1)^{N_{M-1}}) \label{twist}
\end{equation}
(ensuring $\g^M=1$) and
Cartan generators are represented by
$2 \times 2$ $\sigma_3$ submatrices.

In this basis, projection equation \ref{thetaprojection} reduces to the simple condition
\begin{equation}
\rho _k V = \d_k \label{proyeccion}
\end{equation}
where $\rho _k$ is the weight vector associated to the corresponding $\lambda ^k$  representation. 

Similarly we can represent  $\Omega$ Chan-Paton action in terms of
a unitary matrix $\g_{\Omega}$. More generally, the action of
$\Omega g$, $g \in Z_M$, is given by
\begin{equation}
\Omega g : |\Psi, ab ) \rightarrow (\g_{\Omega g, p})_{aa'}
|\Omega g \Psi, b'a' )(\g_{\Omega g, q})_{b'b}^{-1}
\end{equation}
Consistency with the orientifold group multiplication law
implies several constraints on the $\g$ matrices like
\begin{equation}
\g_{\Omega g,p}= \g_{g,p} \g_{\Omega,p}
\end{equation}
or from $(\Omega \theta ^x)^2= \theta ^{2x}$,
\begin{equation}
\g_{\Omega x,p}= \pm \g_{2x,p} \g_{\Omega x,p}^T.
\end{equation}
Cancellation of tadpoles imposes further conditions on the $\g$ matrices.

To summarize this section, notice that had we managed to find a consistent model with known Chan Paton factor $\lambda ^k$, for instance from consistency restrictions on the boundary states and the direct channel amplitudes, phase modded models can then be easily constructed. Open string states are just an invariant subset of the original ones.
Recall that, even if the initial group were non-chiral , as it is the
case if we started from a diagonal ($k$ odd) invariant (leading to  $SO(n)$ or $Sp(n)$ gauge groups), the projection condition on gauge bosons  $\rho _k V = 0$ could lead to unitary groups.
 This works in much the similar way as orbifold invariant states are selected from $SO(32)$ D9-brane Chan-Paton factors in Type I, odd orbifold, compactifications.
Certainly,  not any   projection will be allowed since  tadpole cancellation conditions must still be satisfied.

Interestingly enough, for the  $k$ even case,  $y=0 \mod{ m/2}$ twisted sector will appear. Thus, new states, absent in the starting theory, will be generically present. This signals the presence of new type of branes with open strings stretching between them  as, for instance, D5-branes appear in Type I, even orbifold, compactifications. Extra tadpole cancellation equations will be associated to such states.

\subsection{An index formula for  Gepner Models }

  A simple  index formula can be explicitly written down for Gepner models. This can be achieved by adapting to Gepner case the expressions derived in ~\cite{bm} for Witten index $\Tr (-1)^F$ in generic conformal field theories, for bifundamental representations.
Such an expression could help, in particular, to determine  modular invariants and/or phase moddings $\G$  that could lead to chiral matter content.

We find, for Gepner models (see partition function in \ref{cild}) ,
\beq
I_{\a \b}= \#(n_\a, \bar n_\b)-\#(\bar n_\a,  n_\b) =  \sum_{\g} i e^{ i\frac{\pi}{2} Q_{\g}}   C_{\a\b}^{\g}
\eeq
where $Q_{\g} = -\sum_{i=1}^r \frac{(q_{\g})_i}{m_i}$.

Notice, for example,  that for  a diagonal modular invariant,
chirality vanishes. In fact, such couplings verify
\beqa
C_{\a\b}^{\g} &=& C_{\a\b}^{\g^*}\\
Q_{\g}&=&-Q_{\g^*}
\eeqa
where $\g^*$ denotes the vector with components $(\vec {l}_\g,- \vec {q}_\g)$.

If Gepner  model is further modded  by discrete symmetries, a projector onto invariant states
in the trace $\frac{1}{M} \sum \theta^x$ must be included.

Bifundamental chirality  now reads
\beq
I_{\a,j; \b,i}=\#(n_{\a,j}, \bar n_{\b,i})-\#(\bar n_{\a,j}  n_{\b,i})= \sum_{\g} i e^{ i\frac{\pi}{2} Q_{\g}}   C_{\a\b}^{\g}
\delta_{\frac{\G q_{\g}}{m} +V_j^{\a}-V_i^{\b},0}
\eeq
where $V_j$ is given by (\ref{twist}).
Thus we see that this index is not necessarily  vanishing, due to the presence of the $\delta$ function.

\section{Tadpole cancellation}
We proceed to  write down the amplitudes in the direct and  the transverse channel in order
to study factorization and tadpole cancellation. For the sake of simplicity,
we first consider the case of ($k$ odd) pure Gepner model and then we generalize to
the hybrid case.
The  cylinder amplitude is given by
\begin{equation}
Z_C(it ) = \frac{1}{M}\sum_{x=0}^{M-1}\sum_{\g\a\b}C_{\a\b}^{\g}
\tr \g_{\a,x}\tr \g_{\b,x} e^{-2i\pi \G\frac{q}{m}x}\chi_{\g}(it)
\end{equation}
which is nothing but  the generalization of (\ref{cild}) when
a projection operator $\frac{1}{M}\sum\t ^x$ is included in the trace.
The transverse channel  representation of this amplitude reads
\begin{equation}
\tilde Z_C(il ) = \frac{1}{M}\sum_{x=0}^{M-1}\sum_{\a\b\g\delta}C_{\a\b}^{\g} S_{\g\delta}
\tr \g_{\a,x}\tr \g_{\b,x} \chi_{\delta+2 \G x}(il)
\end{equation}
where we have used (\ref{matrizs}).
Notice that for each fixed $x$, (\ref{da}) is  verified with $n_a \rightarrow \tr \g_{a,x}$. Namely,
\begin{equation}
 C_{\a\b}^{\g} S_{\g\delta}
\tr \g_{\a,x}\tr \g_{\b,x} = (D^{\b}_{\a} \tr \g_{\a,x})^2
\end{equation}
indicating that the transverse amplitude can again be written as a square
\begin{equation}
\tilde Z_C(il ) = \frac{1}{M}\sum_{x=0}^{M-1}\sum_{\a\b}(D^{\b}_{\a} \tr \g_{\a,x})^2 \chi_{\b+2 \G x}(il)
\end{equation}

Finally, the M\"obius strip also contributes at one loop in the open string
sector in the following way
\begin{equation}
Z_M(it +1/2 ) = \frac{1}{M}\sum_{x=0}^{M-1}\sum_{\g\a}M_{\a}^{\g}
\tr (\g_{\Om x,\a}^{-1} \g_{\Om x,\a} ^{T} )e^{-2i\pi \G\frac{q}{m}x}\hat \chi_{\g}(it+1/2)
\end{equation}
where again we introduce the real \textit{hatted} characters.
Thus,  transverse M\"obius strip amplitude reads
\begin{equation}
\tilde Z_M(il +1/2 ) = \frac{1}{M}\sum_{x=0}^{M-1}\sum_{\g\a}2^{\frac{D}{2}}\tilde M_{\a}^{\g}
\tr (\g_{\Om x,\a}^{-1} \g_{\Om x,\a} ^{T} ) \hat \chi_{\g+4\G x}(il+1/2).
\end{equation}
where $\tilde M_{\a}^{\delta} = P_{\delta\g}M_{\a}^{\g}$.

Collecting  contributions from  Klein bottle, cylinder and
M\"obius strip in the transverse channel we have
\begin{eqnarray}\nonumber
\frac{1}{M}\sum_{x=0}^{M-1} \sum_{\a}
 \{ && 
(O^{\a})^2 \chi_{\a}(il) +(D^{\a}_{\b} \tr \g_{\b,x})^2 \chi_{\a}(il)+ \\
& &
2 \times 2^{\frac{D}{2}}\tilde M_{\b}^{\a}\tr (\g_{\Om x,\b}^{-1} \g_{\Om x,\b} ^{T} )  \hat \chi_{\a}(il + \um) \}
\end{eqnarray}
Notice that despite  there are no $\G$ twisted sectors in the direct
channel, projections lead to $x$ twisted sectors in the transverse one.

Factorization (\ref{factori}) for  $l \rightarrow \infty$, amounts to
\begin{equation}
(D^{\a}_{\b} \tr \g_{\b,2x})^2 +
2 \times 2^{\frac{D}{2}}\tilde M_{\b}^{\a}\tr (\g_{\Om x,\b}^{-1} \g_{\Om x,\b} ^{T} ) +
(O^{\a})^2 = \textrm{perfect square}
\end{equation}
Namely, $2^{\frac{D}{2}} \tilde M_{\a}=  D_{\a}O_{\a}$ and
\begin{equation}
\tr (\g_{\Om x,\a}^{-1} \g_{\Om x,\a} ^{T} )= \pm \tr \g_{2x,\a}
\end{equation}
the same type of condition as found in orbifold compactifications \cite{gj}.
Interestingly enough, we see that, with this condition, factorized
amplitudes in the modded theory can be immediately written down from
the unprojected theory by just performing the replacement
\begin{equation}
n_{\a} \rightarrow \tr \g_{\a,x}
\end{equation}

Moreover,  zero charge condition for RR massless fields (\ref{tadpolecg}) is easily found by requiring
\begin{equation}
D^{\a}_a \tr \g_{a,x}+
O^{\a} =0
\label{tadcanproj}
\end{equation}
for  characters $\chi_{\a+2 \G x}$ containing massless RR states.

In particular, for untwisted  $x=0$ transverse sector the original conditions are recovered since, $\tr \g_{a,0}=n_a$.
As mentioned, when diagonal invariants are considered, the number of  tadpole conditions is smaller than with other invariants, and therefore easier to solve.

 Nevertheless after modding
out phase symmetries this number rises and it corresponds to the number of massless twisted transverse characters.

Tadpole condition can be generalized  for   hybrid models $T^{2(3-n)} \times Gepner$ and $Z_N$ modding, odd N, in the following  way
\begin{equation}
D_{\b}(\tr \g_{\a ,x}) + 2O_{\b}  \prod_{i=0}^{3-n} 2 \cos \pi x v_i  =0 \label{tadhyb}
\end{equation}
which arises from transforming to the transverse channel (\ref{kleinorb}) and
the open sector amplitudes
\begin{displaymath}
Z_C(it)= \left \{ \sum_{\alpha \b \g}
\left [\frac{\theta\left[{0 \atop 0}\right]}{\eta^3} \right]^2
\prod_{i=1}^{3-n} \frac{\theta\left[{0 \atop {xv_i}}\right]}{\eta} \frac{-2 \sin{\pi x v_i }\eta}
{\theta\left[{\frac{1}{2} \atop {\frac{1}{2}+2xv_i}}\right]}
C_{\a\b}^{\g} e^{-2i\pi \G x\frac{q}{m}} \chi_{\g}
\Tr \gamma_{x,\a} \Tr \gamma_{x,\b} \right \}_{susy}
\end{displaymath}
\begin{displaymath}
Z_{M}(it+\um)= \left \{ \sum_{\a a}
\left [\frac{ \theta\left[{0 \atop 0}\right]}{\eta^3} \right]^2
\prod_{i=1}^{3-n} \frac{\theta\left[{0 \atop {xv_i}}\right]}{\eta} \frac{-2 \sin{\pi x v_i }\eta}
{\theta\left[{\frac{1}{2} \atop {\frac{1}{2}+2xv_i}}\right]}
M_{a}^{\a}e^{-2i\pi \G x\frac{q}{m}}\hat  \chi_{\a}
\Tr [\gamma^{-1}_{\Omega x,a}\gamma^{T}_{\Omega x,a}] \right \}_{susy}
\end{displaymath}
We should remember that according to the combined action of the orbifold twist and $\Omega$, a sum over quantized momenta or over windings must be included (see ~\cite{afiv}).
To arrive at the tadpole cancellation conditions, we must take the limit $t \rightarrow 0$ in the various traces and next change variable to $l$ appropriately to find the large $l$ behavior of the amplitudes. The final step is to collect all terms with a given volume dependence.

\section{Examples }
In this section we exhibit some explicit examples of $D=4$ chiral models  by following the general steps discussed above.
The situation in which the internal sector is \textit{purely } Gepner is illustrated by considering  phase moddings of $3^5_D$ quintic.
The hybrid situation of a Gepner-orbifold internal sector is exemplified  by considering  orbifolds of  $(1)^6 \times T^2 $ and  $(1)^3 \times T^4 $ internal sector models.
The latter is a peculiar model, in some sense, since the Gepner part is, actually, also a (special) torus. Nevertheless, it is useful not only to illustrate the general method  but to show how models where rank is reduced are obtained .  By closely following Ref ~\cite{aiq2} we show how antibranes  can be included in these constructions.

\subsection{${\bf 3^5_D /\IZ_5}$ }
A consistent open string theory with internal sector given by ${\bf 3^5}$ Gepner model, with a diagonal invariant, was found in \cite{aaln} from where we borrow results and notation (see also \cite{blumen} and \cite{bhhw}).
The KB partition function can be written as
\begin{equation}
{\cal Z}_{K}(it)=\frac 12 \frac{1}{5}\left [(\chi _{I}(it)+\chi
_{II}(it))^{5}\right ]^{susy}
\end{equation}
where
\begin{eqnarray}
\chi _{I} &=&\chi _{(0,0)}+\chi _{(3,-3)}+\chi _{(3,-1)}+\chi _{(3,1)}+\chi
_{(3,3)}  \nonumber \\
\chi _{II} &=&\chi _{(2,0)}+\chi _{(2,2)}+\chi _{(1,-1)}+\chi _{(1,1)}+\chi
_{(2,-2)}
\end{eqnarray}
which reads, in the transverse channel
\begin{equation}
\tilde{\cal Z}_{K}(il)=
\frac 12 2^{4}\sqrt[4]{5}\left [(\kappa^{\frac{3}{2}}\tilde{\chi}%
_{(0,0)}(il)+\kappa^{-\frac{3}{2}}\tilde{\chi}_{(2,0)}(il))^{5}\right
]^{susy}
\label{kbtrans35}
\end{equation}
where $\kappa\equiv \frac{1}{2}(1+\sqrt{5})$.
The partition function in the transverse channel can be written in
terms of $\tilde{\chi}_{(0,0)}(il)^{1-\gamma _{i}}
\tilde{\chi}_{(2,0)}^{\gamma _{i}}$  (where the exponents indicate the
number of times each factor appears, regardless of order).
 Each term is encoded in a
5 component vector (one for each theory) $\vec\gamma$ taking values
$0$ or 1 (which corresponds to a state belonging to group I or II in the direct channel, respectively)
For instance, by rewriting  \ref{kbtrans35} as,
\begin{equation}
\tilde{\cal Z}_{K}(il)==\frac
12 {\cal O}^2_{\vec{\gamma}}\left
[\prod\limits_{i=1}^{5}(\tilde{\chi}%
_{(0,0)}(il))^{1-\gamma _{i}}(\tilde{\chi}_{(2,0)}(il))^{\gamma
_{i}}\right ]^{susy}
\label{cil35}
\end{equation}
we find that the only non vanishing coefficients are
\begin{equation}
{\cal O}_{\vec{0}} =2^{4}5^{\frac18}\kappa^{\frac{15}{2}}  \qquad ; \qquad
{\cal O}_{\vec{1}} =2^{4}5^{\frac18}\kappa^{-\frac{15}{2}}  \label{Tadpole1}
\end{equation}
where $\vec 0 \equiv (0,0,0,0,0)$ and $\vec 1 \equiv (1,1,1,1,1)$.

Similarly ${\cal D}_{\vec{\gamma}}$ and ${\cal M}_{\vec{\gamma}}$
coefficients are defined for the cylinder and MS
amplitudes. Consistency is ensured for \cite{aaln}
\begin{eqnarray}
{\cal D}^2_{\vec{\gamma}} &=&\frac{5^{\frac14}}{\kappa^{5/2}}\frac{\left(
\sum\limits_{\vec{\delta}%
}\kappa^{(\vec{\gamma}-\vec{\delta})^{2}}(-1)^{\vec{\gamma}.
\vec{\delta}}n^{\vec{%
\delta}}\right) ^{2}}{\kappa^{(\vec{\gamma})^{2}}}
\end{eqnarray}
and
\begin{equation}
\tilde{\cal
M}_{\vec{\gamma}}=\tilde{\cal
D}_{\vec{\gamma}}\tilde{\cal
O}_{\vec{\gamma}}=-\sum\limits_{\vec{\delta}}\sqrt[4]{5}(-1)^{(\vec{%
\gamma})^{2}}\kappa^{(\vec{\gamma}-\vec{\delta})^{2}}(-1)^{\vec{\gamma}.\vec{%
\delta}}\kappa^{\frac 52 - 2(\vec{\gamma})^{2}}n_{\vec{\delta}} .
\end{equation}
Tadpole cancellation equations are thus
\begin{eqnarray}
{\cal D}_{\vec{0}}+{\cal O}_{\vec{0}} &=&0\\
{\cal D}_{\vec{1}} +{\cal O}_{\vec{1}} &=&0
\end{eqnarray}
which read,
\begin{eqnarray}
N_{0}+N_{2}+N_{3}+2N_{4}+3N_{5} &=&12  \nonumber \\
N_{1}+N_{2}+2N_{3}+3N_{4}+5N_{5} &=&20
\label{Tadpoles35}
\end{eqnarray}
where $N_{i}=\sum\limits_{\vec{\gamma}\;/\;\left| \vec{\gamma}\right|
=i}n_{\vec{\gamma}}\quad$
(e.g.,
$n_{4}=n_{(1,1,1,1,0)}
+n_{(1,1,1,0,1)}+n_{(1,1,0,1,1)}+n_{(1,0,1,1,1)}+n_{(0,1,1,1,1)}
$).
By studying the direct channel expressions the open string spectrum
can be found\cite{aaln}. The gauge group is of the form $\prod_{i=0}^5
SO(n_i)$ with matter states transforming in antisymmetric, symmetric
or bifundamental representations.
For instance, with
$N_{0}=n_{(0,0,0,0,0)};N_{1}=n_{(1,0,0,0,0)};N_{2}=n_{(1,1,0,0,0)}$ (and all other entries vanishing)
the following massless spectrum is found
\[ {\renewcommand{\arraystretch}{1.3}
\begin{tabular}{|c|c|c|c|}
\hline
S-T & Internal & mult. & irrep.\\
\hline
v & $(0,0)^5$                        & 1 &
$SO(N_0) \otimes
SO(N_1) \otimes SO(N_2)$\\
s & $(2,2) \underline{(3,3)(0,0)^3}$ & 4 &
 $(1,\Ysymm,1) + (1,1,\Ysymm) + (N_0,N_1,1)$\\
s & $ (3,3)(2,2)(0,0)^3 $             & 1 &
 $(1,1,\Ysymm) + (1,N_1,N_2)$\\
s & $(0,0)(2,2)\underline{(0,0)^2(3,3)}$ & 3 &
 $(1,1,\Ysymm) + (1,N_1,N_2)$\\
s & $(1,1)^2 \underline{(0,0)^2(3,3)}$ & 3 &
 $(1,1,\Yasymm) + (N_0,1,N_2) + (1,N_1,N_2)$\\[.5ex]
\hline
\end{tabular}
}\]
with
\begin{equation}
N_{0} =12-N \quad ; \quad N_{1}=20-N  \quad ; \quad N_{2} =N
\label{35example}
\end{equation}
$v,s$ indicate that such states are vector or chiral scalar superfields, respectively.

Phase symmetries of the $3^5$ allow for $124$ different independent moddings. Moreover, more than one modding could be simultaneously performed.
By embedding such moddings as twists on D-branes chiral models can be obtained.
In order to illustrate the procedure let us consider the  simple situation when
$N=0$.  Thus, our starting point is a
\begin{eqnarray} \nonumber
&&SO(12) \otimes SO(20) \\
&& 4[(1,\Ysymm) + (12,20)]
\end{eqnarray}
where, as can be read from the table, the multiplicity comes from possible permutations in
$$(2,2) {(3,3)(0,0)^3}+(2,2) {(0,0)(3,3)(0,0)^2}+(2,2) {(0,0)^2(3,3)(0,0)}+(2,2) {(0,0)^3(3,3)}
$$
We choose to perform the modding  $\Gamma = (0,2,-1,-1,0)$ and to embed it as the generic  Chan-Paton twist
\begin{equation}
V=\frac{1}{5}(0^{n_0},1^{n_1},2^{n_2};0^{m_0},1^{m_1},2^{m_2})
\end{equation}
with
\begin{eqnarray}
\oh N_0&=&{n_0}+{n_1}+{n_2}=6\\ \nonumber
\oh N_1&=&{m_0}+{m_1}+{m_2}=10
\end{eqnarray}
Spectrum is obtained by projecting the above states according to eq.\ref{thetaprojection}. Here
$\delta =\frac{\G .q}{5}= 0,-\frac{1}{5},\frac{2}{5},\frac{2}{5},0$ for the gauge bosons and the four respective massless matter states while, for instance,
\begin{eqnarray}
\rho _{(Adj,1)}&=&(\underline{\pm 1,\pm 1,0,\dots,0};0\dots,0)\\
\rho _{(1,\Ysymm)}&=&(0,\dots,0;(\underline{\pm 1, \pm 1,0,\dots,0}))+
(0,\dots,0;\underline{\pm 2,\dots,0})\\
\rho _{(12,20)}&=&(\underline{\pm 1,0,\dots,0};\underline{\pm 1,0,\dots,0})
\end{eqnarray}
where the first (second) 6 (10) entries correspond to $SO(12)$ ($SO(20)$) weight vectors.
Thus, we see, for instance  that the original gauge group breaks to
$SO(2n_0)\times U(n_1)\times U(n_2)\times SO(2m_0) \times U(m_1)\times U(m_2)$.
Matter states can be easily computed. Spectrum is generically chiral but anomalous for arbitrary values of $n$'s and $m$'s. In fact, strong restrictions are imposed by tadpole cancellation.

As showed above, tadpole cancellation equations for the projected theory can be easily obtained from the unprojected theory (see \ref{tadcanproj}.)
We just have to replace $n_a \rightarrow \tr \g_{a,x}$ in the transverse channel expressions for the corresponding $x$ twisted characters.
Namely,
\begin{eqnarray}\nonumber
N_0& \rightarrow & \tr \g_{0,x}=2n_0+2 n_1 \cos{\frac25\pi x}+2n_2 \cos{\frac45\pi x}\\
N_1& \rightarrow & \tr \g_{1,x}=2 m_0+2m_1 \cos{\frac25\pi x}+2 m_2 \cos{\frac45\pi x}
\label{untwist}
\end{eqnarray}

Tadpole cancellation equations thus read
\begin{equation}
( D^{\a}_a \tr \g_{a,x}+
O^{\a}  )^2 \chi_{\a+2 \G x}(il)=0
\quad \quad (l \rightarrow \infty)
\end{equation}
for all $x$ twisted states such that $\vec \alpha + 2 \Gamma x$ contains an RR massless state.
For the example at  hand such states are given in table \ref{xmass} below
\begin{table}[!ht]
\begin{center}
\begin{tabular}{|ccccc|c|ccccc|}
\hline
\multicolumn{5}{|c|}{$\vec \alpha$} & x &
\multicolumn{5}{|c|}{$\vec \alpha + 2 \Gamma x$ (massless)} \\
\hline
(0,0)&(0,0)&(0,0)&(0,0)&(0,0)&0&
(0,0)&(0,0)&(0,0)&(0,0)&(0,0)\\
(2,0)&(2,0)&(2,0)&(2,0)&(2,0)&0&
(1,-1)&(1,-1)&(1,-1)&(1,-1)&(1,-1)\\
(0,0)&(2,0)&(2,0)&(2,0)&(0,0)&1&
(0,0)&(1,-1)&(2,-2)&(2,-2)&(0,0)\\
(2,0)&(0,0)&(0,0)&(0,0)&(2,0)&2&
(1,-1)&(3,-3)&(0,0)&(0,0)&(1,-1)\\
(0,0)&(0,0)&(2,0)&(2,0)&(0,0)&3&
(0,0)&(3,-3)&(1,-1)&(1,-1)&(0,0)\\
(2,0)&(2,0)&(0,0)&(0,0)&(2,0)&4&
(2,-2)&(1,-1)&(0,0)&(0,0)&(2,-2)\\
\hline

\end{tabular}
\end{center}
\caption{Massless twisted states $\vec \alpha + 2 \Gamma x$  } \label{xmass}
\end{table}

and lead to the equations
\begin{eqnarray*}
44 + 20 \sqrt{5} &=& \pm (2 \;\tr \g_{0,0} + \;\tr \g_{1,0} + \sqrt{5} \;\tr \g_{1,0})\\
 8              &=& \pm (11 \;\tr \g_{0,0} + 5 \sqrt{5} \;\tr \g_{0,0} - 7 \;\tr \g_{1,0} - 3\sqrt{5} n_1)\\
12 + 4 \sqrt{5} &=& \pm (4 \;\tr \g_{0,2} + 2 \sqrt{5} \;\tr \g_{0,2} + 7 \;\tr \g_{1,2} + 3 \sqrt{5} \;\tr \g_{1,2})\\
12 + 4 \sqrt{5} &=& \pm (4 \;\tr \g_{0,2} + 2 \sqrt{5} \;\tr \g_{0,2} - 3 \;\tr \g_{1,2} - \sqrt{5} \;\tr \g_{1,2})\\
16 + 8 \sqrt{5} &=& \pm (3 \;\tr \g_{0,4} + \sqrt{5} \;\tr \g_{0,4} + 4 \;\tr \g_{1,4} + 2 \sqrt{5} \;\tr \g_{1,4})\\
16 + 8 \sqrt{5} &=& \pm (3 \;\tr \g_{0,4} + \sqrt{5} \;\tr \g_{0,4} - \;\tr \g_{1,4} - \sqrt{5} \;\tr \g_{1,4})
\end{eqnarray*}

Sign freedom is due to the fact that both signs lead to square completion. The first two equations are the original, untwisted ones (\ref{untwist}) fixing the total group ranks. The extra, \textit{x twisted}, equations can be easily checked to  be the conditions to be satisfied for the theory to be \textit{anomaly free}.
The solution to these tadpole equations is unique in this case (it corresponds to sign selection$\{+, +, -, +, -, -\}$), namely
\begin{center}
\begin{tabular}{c}
$n_{0}=2\quad{}n_{1}=4\quad{}n_{2}=0$\\
$m_{0}=2\quad{}m_{1}=4\quad{}m_{2}=4$\\
\end{tabular}
\end{center}
leaving an
$SO(4)\times U(4)\times SO(4)\times U(4)\times U(4)$ gauge group with chiral matter content
{\footnotesize{ \begin{eqnarray}\nonumber
&(1,1;4,4,1)+(1,1;1,{\bar 4},4) +(1,1;1,1,\overline{10})+(1,{\bar 4};1,1,4)+
(1, 4;1,{\bar 4},1)+(4,1;1,4,1)+\\
&2[1,1;1,10,1)+(1,1;1,{\bar 4},{\bar 4})+(1,1;4,1,4)+(1,{\bar 4};1,{\bar 4},1)+(4,1;1,1,4)+(1,4;4,1,1)]\nonumber
\end{eqnarray}}}
We thus see that unitary groups with chiral matter can be easily obtained.

A biased search by considering simultaneous  moddings should be performed in order to look, for instance,  for models closer to the Standard model or some extension of it. We are not attempting this search here.

Nevertheless, we have checked the  spectrum for the 124 independent moddings. For the $SO(N_0)\times SO(N_1)\times SO(N_2)$ case of (\ref{35example}) we found that sixteen of them  lead to inconsistent models where tadpoles can not be cancelled.
For some of the allowed moddings more than one solution to tadpole equations exist.

The following 36 moddings lead only to non-chiral models:
\[
\G=(0,0,{\underline{0,1,-1}}),(0,0,{\underline{0,2,-2}}),
\pm(0,1,{\underline{-1,2,-2}}),\pm(0,2,{\underline{1,-1,-2}})
\]
The other 72 all lead to at least a solution with chiral spectra.
Unitary groups with high ranks, up to $U(10)$ are found.  However,
only groups of up to rank four have chiral spectra (see also \cite{bw}).
Clearly further projections leading to further breaking might lead to
other possibilities.

We have checked, in some examples, that twisted tadpole cancellation
does coincide with anomaly cancellation conditions.

\subsection{${\bf (1^3 \times T^4) /\IZ_3}$}
An open string theory with internal sector given by $(1)^3$ Gepner model, with a diagonal invariant, was
found in ~\cite{aaln} . In this case the KB, MS and C partition functions can be written in the transverse channel as
\beq
\tilde Z_K + \tilde Z_M + \tilde Z_C = 2^8 {\tilde \chi}^{susy}_{(0,0)^3}(il) - 2 \times 2^4 {\hat {\tilde \chi}}^{susy}_{(0,0)^3}(il+\um)
+n^2{\tilde \chi}^{susy}_{(0,0)^3}(il).
\eeq
Since $\chi^{susy}_{(0,0)^3}$ contains massless RR states, tadpole cancellation equation is thus $n-16 =0$. By studying the direct channel expressions the open
string spectrum can be found. Gauge group is Sp(n) with massive matter states transforming
in antisymmetric or symmetric representation. By compactifying once more on a $T^4$ torus,
we obtain a four-dimensional string theory with open sector massless  spectrum shown in table \ref{una}.

\begin{table}[!ht]
\begin{center}
\begin{tabular}{|c|c|c|}
\hline
Space-time & Internal & Irrep. \\
\hline
v & $(0,0)(0,0,0)^3$ & Sp(16)\\
\hline
s &{$({1, 0})$} $(0,0,0)^3$ & \Ysymm\\
\hline
s &{$({0, 1})$} $(0,0,0)^3$ & \Ysymm\\
\hline
 s & $(0,0)(1,-1,0)^3$ &\Ysymm\\
\hline
\end{tabular}
\end{center}
\caption{Open sector massless spectrum} \label{una}
\end{table}

These states make up  a $Sp(16)$ vector multiplet and 3 chiral superfields transforming in $ (\Ysymm$) .
Phase symmetries of the $(1)^3 \times T^4$ allow for two different independent moddings
\beqa
\G&=&(1,-1,0) \quad v=(0,\ut,-\ut)\\
\G&=&(1,1,0) \quad v=(0,\ut,\ut) \label{mod2}
\eeqa
though the first one leads to N=2 supersymmetries in space-time and hence to non-chiral models.
By embedding the second modding as twist on D-branes chiral models can be obtained.
Consequently, we choose to perform the modding (\ref{mod2})
and to embed it as the generic Chan-Paton twist
\beq
V= \ut(0^{n_0}, 1^{n_1}) \label{twist1}
\eeq
with $n_0 + n_1=8$. Spectrum is obtained by projecting above states according to eq. (\ref{proyeccion}). Here
$\d =\frac{\G q}{m}=0,\ut,\ut,\ut$ for the gauge bosons and three respective massless states.
Thus, from (\ref{twist1}) and
(\ref{proyeccion}) we obtain that the original gauge group breaks to $Sp(2n_0) \times U(n_1)$. Matter states can
be easily computed and lead to a $3[(2n_0,n_1) + (1,\overline{\Ysymm})]$ chiral representation. Spectrum
is chiral but anomalous for arbitrary values of $n_0, n_1$. However, strong restrictions  imposed by tadpole
cancellation will ensure anomaly-free spectrum.

As we have shown, above tadpole cancellation equations for the projected theory can be easily obtained
 from  the transverse channel expressions. Tadpole cancellation
equations given in  (\ref{tadhyb}) thus read
\beq
\Tr \g_x - 2 (2 \cos{\frac{\pi x}{3} })^2=0
\eeq
for all $x$  such that $\chi_{\a +2\G x}$ contains a RR massless state. Such states are given
in table \ref{massRR} below.

\begin{table}[!ht]
\begin{center}
{
\renewcommand{\arraystretch}{1.3}
\begin{tabular}{|c|c|}
\hline
x & $\a + 2 \G x$ = St $\times$ Gepner   \\
\hline
0 &$(-\um,-\um,-\um)$;{$(0,1,1)(0,1,1)(0,1,1)$} + others\\
\hline
1 &$(-\um,-\um,-\um)$;\underline{$(0,-1,-1)(0,-1,-1)(0,1,1)$}\\
\hline
2 &$(\um,\um,\um)$;\underline{$(0,1,1)(0,1,1)(0,-1,-1)$} \\
\hline
\end{tabular}
}
\end{center}
\caption{RR massless states in transverse channel} \label{massRR}
\end{table}

They lead to the equations
\beqa
n_0 + n_1 &=&8 \\ 2n_0 - n_1 &=& 4.
\eeqa
The first equation is the original, untwisted one. The extra, twisted, equation is just the condition for the theory to be anomaly free.

The solution to these tadpole equations is unique, namely $n_0=n_1=4$ leaving a $Sp(8)\times U(4)$ gauge
group with chiral matter content $3[(8,4)+(1,\overline \Ysymm)]$.

Similarly to the $3^5$ Gepner model we see that unitary groups with chiral matter can be easily obtained
in the hybrid models. Moreover, this example leads to
 three matter generations. It is interesting to compare this computation with an orbifold like case.
For instance, in a type-IIB orientifold on $T^6/Z_3$ 
the massless spectrum reads
\begin{eqnarray}
Vector && SO(2n_0) \times U(n_1) \\
Chiral && 3[(2n_0, n_1) + (1, {\overline \Yasymm})]
\end{eqnarray}
and the tadpole cancellation conditions are
\begin{eqnarray}
 n_0 + n_1 &=& 16 \\
 2n_0 -  n_1 &=&-4.
\end{eqnarray}
Thus, compared to our case, we see that they are similar if we make the following
replacement: $SO(n) \rightarrow  Sp(n)$ and $\Yasymm \rightarrow \Ysymm$. Besides, notice that
the Gepner model leads to a rank reduction which can be explained from the presence of a NSNS antisymmetric field.

\subsection{{\bf {Non-supersymmetric standard like models}}}

In this section we  consider a non-supersymmetric variation of the $(1)^3 \times T^4$ model discussed in the previous
section. We show that by  finding a specific form of the twist matrices,  tadpole cancellation might require
the presence of parallel branes and antibranes on the torus $T^{2(3-n)}$ sector.
In addition, adding  Wilson lines, for example one wrapped in the $e_1$ direction in the first
complex plane,  will enormously increase the freedom to construct phenomenologically interesting
three generations models.

Let us begin with a $(1)^3 \times T^4$ model and add open strings with the following boundary conditions
\beqa
n^a \partial_a X^{\mu} &=& 0 ,\quad p=0,...,7\\
X^{i} &\in& (1)^3 , \quad i=8,9
\eeqa
and
\beqa
n^a \partial_a X^{\mu} &=& 0 ,\quad p=0,...,3\\
X^i & =&0, \quad i=4,5,6,7 \\
X^{i} &\in& (1)^3 ,\quad i=8,9
\eeqa
Both groups
of conditions define that open string ends  live, respectively, in what  we have generically called,  DQ-branes and DP-branes. 
In order to cancel untwisted tadpoles the orientifold action requires the introduction of
16 DQ-branes and a zero net number of DP-branes. However,  it is also possible to  include DP-branes if an equal number of D$\bar P$-antibranes is  introduced (wrapping, for instance, the third complex plane). This leads to new (non-supersymmetric) consistent models. This example is closely related to the non-supersymmetric  $Z^3$  orientifold of ~\cite{au}.
Branes and antibranes annihilation can be prevented by considering models with branes and antibranes stuck at different fixed points in $T^4$. The tadpole cancellation conditions read
\begin{equation}
\Tr  \g_Q +3(\Tr \g_{P,L} - \Tr \g_{\bar P,L})=4
\end{equation}
for any of the nine fixed points $L$ in the two first complex planes. Also the number of DP-branes and
D$\bar P$-antibranes must be the same. Notice that choosing $\Tr  \g_Q \ne 4$ inevitably demands the
presence of DP-branes and/or D$\bar P$-antibranes at all fixed points. We also include a Wilson line
$W$ wrapping along direction $e_1$ in the first complex plane which modifies the tadpole cancellation
equations
\begin{equation}
\Tr W^k \g_Q +3(\Tr \g_{P,L} - \Tr \g_{\bar P,L})=4 \qquad k=0,1,2
\end{equation}

We choose to perform the modding (\ref{mod2}) and to embed it as the generic Chan-Paton twists
\beqa
V_Q  & =& \frac{1}{3}(0^{N_0},0 ^{N_1},1^{N_2}, 1^{N_3},1^{N_4})\\
 W &=&\frac{1}{3}(0^{N_0},1 ^{N_1},0^{N_2}, 1^{N_3},2^{N_4})\\
V_P &=& \ut (0^{m_L}, 1^{n_L})\\
V_{\bar P} &=& \ut (0^{p_L}, 1^{q_L})
\eeqa
with $N_0 + N_1+N_2+N_3+N_4 =8$. Spectrum is obtained by projecting states as above, but now it contains
new states coming from different open string sectors, namely, QQ, PP, QP, Q$\bar P$, etc. For instance,
bosonic massless states in the $QP$ sector will be given by
\beq
(0,\um,\um)(0,0,0)^3
\eeq
due to the fact that there are mixed DN boundary conditions  in the $X^p , p=4,5,6,7$ coordinates.
Thus, we see, for example, that the original gauge group in the QQ sector breaks to
\begin{equation}
Sp(2N_0)\times \prod_{s=1}^{4} U(N_s).
\end{equation}

Matter states can be easily computed following the steps used in the $Z_3$ orientifold of ~\cite{aiq2}. We leave
this long computations to the Appendix. We illustrate the results in some interesting examples below.
 Considering three different actions of the twist on the Chan-Paton factors $\{N_i\}$
we lead to the models in table \ref{standard}.

\begin{table}[!ht]
\begin{center}
\begin{tabular}{|c|c|}
\hline
Observable Group & Chiral Matter (3 families)\\
\hline
$SU(3)\times SU(2)\times U(1)_Y$ &  $  (3,1)_{\frac{2}{3}}+(\bar 3, 2)_{-\frac{1}{6}}+  (3,1)_{-\ut}$\\
 &$   (1,1)_{-1}+ (1,2)_{+\um}$ \\
\hline
 $SU(3)\times SU(2)_L \times SU(2)_R \times U(1)_{B-L} $ &
 $ (3,2,1)_{-\frac{1}{3}}+(\bar 3,1,2)_{+\ut}+( 1, 2,2)_{0}$\\
& $ (1,2,1)_{-1}+(1,1,2)_{+1} $\\
\hline
$SU(4)\times SU(2)_L \times SU(2)_R \times U(1)_{B-L}$ &
 $(4,2,1)_{-\frac{1}{2}}+(\bar 4,1,2)_{+\um}+( 1, 2,2)_{0}$\\
 & $ 2(1,1,2)_{-1} + 2(1,2,1)_{+1}$\\
\hline
\end{tabular}
\end{center}
\caption{Standard-like models} \label{standard}
\end{table}

We see that Standard like model, Left-Right symmetric models or Pati-Salam models with three generations can be easily constructed.
Interestingly enough, the allowed group,  rank 8, is fulfilled by the Pati-Salam group in the $QQ$ sector.

In addition to above content  non-chiral matter transforming under observable gauge group representations and
matter in some hidden gauge sector  generically appear. Following the general procedure of ~\cite{aiq2} we have identified
one anomaly free combination of the U(1) factors which works as hypercharge.

Recall that while open sector is non supersymmetric, closed sector  has $N=1$ susy, ensured by \ref{forsusy},
and no closed tachyon is present.

\subsection{$( Gepner \,\, model)^{c=6} \times \IT^{2}$}

 Cases in which the internal sector is an orbifold of $c=6$ Gepner models times a two torus can be also considered.
A very simple, toy example,  is provided by starting with  $(1)^6_D$ model of ~\cite{aaln}, and then modding out phase symmetries encoded in
$v=(0,\frac{2}{3})$ and $\G=(-\ut,-\ut,0,0,0,0)$.  $U(4)$ gauge group with matter in $\bf 6$ or $\bf{\bar 6}$ is obtained.
 At first sight, DP branes  could be introduced, as we did in previous $(1)^3 \times T^4$ example, and look for higher rank non-susy extensions by addition of antibranes. However, we find that PP and QQ sectors decouple in this case, and hence do not lead
new massless states.
Further investigation reveals that configurations with DP-branes can exist whenever the Gepner model part contains states with level $k$ even. Interesting non-susy models could be obtained in such cases. We will not analyse them here.

 \section{Summary and outlook}
In this paper we have discussed the construction of Type IIB orientifold models where the internal ``compactified" sector is obtained as a discrete $\IZ_N$ like  symmetry projection of
${( Gepner \,\, model)^{c=3n} \times \IT^{2(3-n)}}$
 space.
 The projection is realized as the combined action of a phase symmetry of Gepner sector and rotation of torus lattice.
Such symmetry action is embedded as a twist on Chan-Paton factors in open sector and leads to  restrictions on them.
By using a Cartan-Weyl basis such restrictions become projections on weight vectors which are very easy to handle.
Generically unitary groups with chiral matter representations are found. We have presented an index formula which allows for a further control on the number of chiral representations.

One interesting outcome of this hybrid construction is that,  presence of parallel branes on the torus $\times \IT^{2(3-n)}$ sector, should allow for lowering the string scale.

A possible strategy, used in our paper, is to start with simple modular invariants for Gepner models, for instance a diagonal one, leading to a small number of tadpole equations.
Once a consistent solution is found, we further project by the corresponding symmetry in order to obtain unitary groups, chiral matter etc. This works in a rather similar way as orbifolds projections of $SO(32)$ group of  Type I string theory.
Tadpole equations are easy to formulate once the starting theory is known. 
An alternative, somewhat opposite,  application of phase moddings is to use them to reduce the number of tadpole equations to solve \cite{bw}.

In this article we have concentrated in some simple  examples in order to illustrate the method rather to attempt a systematic search for phenomenologically interesting models.
An important advantage of the procedure is that several results can be obtained analytically.

Non supersymmetric, tachyon free, open string models were constructed by introducing antibranes, following \cite{au,aiq2}.
Tadpole equations prevent them to annihilate. However, the issue of moduli stabilization remains as an open problem (see for instance \cite{qs}).
We have shown, in the models considered, that anomaly cancellation is ensured by tadpole cancellation.

\centerline{\bf Acknowledgments} We are grateful to L.E. Iba\~nez, F. Quevedo and A. Uranga  for stimulating  discussions and
suggestions. G.A. work is partially supported by PIP 02658 and   ANPCyT grant. E.A. work is supported by Fundaci\'on Antorchas.

\newpage
\section{Appendix}

\subsection{ Non-supersymmetric $(1)^3 \times T^4$ spectrum}

Let us considerer the following twist matrices
\begin{eqnarray}
\tilde \g_Q &=& diag(1_{N_0},1_{N_1},\a 1_{N_2},\a 1_{N_3}, \a 1_{N_{4}}) \\
\tilde \g_{P,i,a}&=& diag(1_{2m^i_a}, \a 1_{n^i_a}, \a^2 1_{n^i_a})\\
\tilde \g_{\bar P,j,a}&=& diag(1_{2p^j_a}, \a 1_{q^j_a}, \a^2 1_{q^j_a})
\end{eqnarray}
with $\a = e^{2\pi i /3}$ and $N_0 + N_1+N_2+N_3+N_4 =8$
 and where nine orbifold fixed points were  labelled as $(a,i), a,i=0,1,2$.
Since the number of branes and antibranes must be the same we must have
\begin{equation}
\sum_{ia}(n^i_a+ m^i_a)=\sum_{bj}(p^j_b+q^j_b)
\end{equation}

We also include a Wilson line $W = (\tilde W, \tilde W^*)$ wrapping along
direction $e_1$ in the first complex plane
\begin{equation}
W= diag(1_{N_0},\a 1 _{N_1}, 1_{N_2}, \a 1_{N_3},\a^2 1_{N_4}) \qquad
\end{equation}
with $\a = e^{2\pi i /3}$.

The tadpole cancellation conditions
when a Wilson line is turned on
read
\begin{equation}
\Tr W^k \g_Q +3(\Tr \g_{P,L} - \Tr \g_{\bar P,L})=4 \qquad k=0,1,2
\end{equation}
for any of the nine fixed points in the two first complex planes. Also the
number of branes and antibranes must be the same.

Using the explicit expressions for twist matrices the tadpole equations read $(j \ne i)$:
\begin{eqnarray}
n_0^i- 2m_0^i =& -q_0^j +2 p ^j_0 &= 4 -N_2 - N_3 -N_4 \\
n_1^i -2m_1^i =& -q_1^j +2 p ^j_1 &= 4 -N_2 - N_3 -N_1 \\
n_2^i- 2m_2^i =& -q_2^j +2 p ^j_2 &= 4 -N_2 - N_1 -N_4
\end{eqnarray}
where we have used that $N_0 + N_1+N_2+N_3+N_4 =8$.
The total gauge group is (when all branes are at fixed points) thus
\begin{equation}
Sp(2N_0)\times \prod_{s=1}^4 U(N_s) \times \prod_{a,i,j \ne i}^2 [SO(2m^i_a)\times U(n^i_a)]\times
[SO(2p^j_a)\times U(q^j_a)]
\end{equation}
 The fermionic spectrum
which is supersymmetric on the branes is given by
\begin{equation}
QQ: \qquad 3 [ (2N_0,N_2)+(\bar N_1,N_3)+(N_1,N_4)+(\bar N_3,\bar N_4)+ \Ysymm_{U_{{N_2}}}]
\end{equation}
\begin{equation}
PP_{L_a}: \qquad 3(2m,n)+2(1,\Yasymm)+(1,\Ysymm)
\end{equation}
\begin{eqnarray*}
QP_{L_0}: fermions - &:&\\
& & (2N_0,n_0^i) +(\bar N_1,n_0^i)+(N_1,n_0^i)+(\bar N_2,\bar n_0^i)+(\bar N_3,\bar n_0^i) \\
 &+&(\bar N_4,\bar n_0^i)+(N_2,2m_0^i)+(N_3,2m_0^i)
+(N_4,2m_0^i)
\end{eqnarray*}
\begin{eqnarray*}
QP_{L_1}: fermions - &:&\\
& & (2N_0,n_1^i) +(\bar N_1,\bar n_1^i)+(\bar N_2,\bar n_1^i)+( N_3,\bar n_1^i)+(\bar N_4, n_1^i) \\
 &+&( N_4, n_1^i)+(N_1,2m_1^i)+(N_2,2m_1^i)
+(\bar N_3,2m_1^i)
\end{eqnarray*}
\begin{eqnarray*}
QP_{L_2}: fermions - &:&\\
& & (2N_0,n_2^i) +( N_1,\bar n_2^i)+(\bar N_2,\bar n_2^i)+( N_3, n_2^i)+( \bar N_3, n_2^i) \\
 &+&( N_4, \bar n_2^i)+(\bar N_1,2m_2^i)+(N_2,2m_2^i)
+(\bar N_4,2m_2^i)
\end{eqnarray*}
In an analogous way we can also compute the non-supersymmetric massless spectrum
for the anti-branes sectors. We find:
\begin{eqnarray*}
\bar P Q_{L_0}: fermions + &:&\\
& & (2N_0,q_0^i) +(\bar N_1,q_0^i)+(N_1,q_0^i)+(\bar N_2,\bar q_0^i)+(\bar N_3,\bar q_0^i) \\
 &+&(\bar N_4,\bar q_0^i)+(N_2,2p_0^i)+(N_3,2p_0^i)
+(N_4,2p_0^i)\\
scalars &:& \\
& & (2N_0,2p_0^j) +(\bar N_1,2p_0^j)+(N_1,2p_0^j) \\
& & +[(\bar N_2,q_0^j)+(\bar N_3,q_0^j)+(\bar N_4,q_0^j) + h.c]
\end{eqnarray*}
\begin{eqnarray*}
\bar P Q_{L_1}: fermions + &:&\\
& & (2N_0,q_1^i) +(\bar N_1,\bar q_1^i)+(\bar N_2,\bar q_1^i)+( N_3,\bar q_1^i)+(\bar N_4, q_1^i) \\
 &+&( N_4, p_1^i)+(N_1,2p_1^i)+(N_2,2p_1^i)
+(\bar N_3,2p_1^i)\\
scalars &:& \\
& & (2N_0,2p_1^j) +(\bar N_4,2p_1^j)+(N_4,2p_1^j) \\
& & +[(\bar N_1,q_1^j)+(\bar N_2,q_1^j)+(\bar N_3,q_1^j) + h.c]
\end{eqnarray*}
\begin{eqnarray*}
\bar P Q_{L_2}: fermions + &:&\\
& & (2N_0,q_2^i) +( N_1,\bar q_2^i)+(\bar N_2,\bar q_2^i)+( N_3, q_2^i)+( \bar N_3, q_2^i) \\
 &+&( N_4, \bar q_2^i)+(\bar N_1,2p_2^i)+(N_2,2p_2^i)
+(\bar N_4,2p_2^i)\\
scalars &:& \\
& & (2N_0,2p_2^j) +(N_3,2p_2^j)+(\bar N_3,2p_2^j) \\
& & +[( N_1,q_2^j)+( N_2,q_2^j)+( N_4,q_2^j) + h.c]
\end{eqnarray*}

\end{document}